\documentclass[aps,prb,superscriptaddress,reprint,showpacs,twocolumn,
preprintnumbers,amsmath,amssymb,floatfix]{revtex4-1}
\usepackage{graphicx}
\usepackage{epstopdf}
\usepackage{color}
\usepackage[latin1]{inputenc}
\usepackage{amssymb}
\usepackage{bm}
\usepackage{stmaryrd}
\usepackage{wasysym}
\usepackage{subfigure}

\begin{document}
\title{Enhancement of low-frequency fluctuations and superconductivity breakdown in
Mn-doped La$_{1-y}$Y$_{y}$FeAsO$_{0.89}$F$_{0.11}$
superconductors}
\author{F. Hammerath}
\affiliation{Department of Physics, University of Pavia-CNISM,
I-27100 Pavia, Italy} \affiliation{Leibniz-Institut f\"ur
Festk\"orper- und Werkstoffforschung (IFW) Dresden, 01171 Dresden,
Germany} \affiliation{Institute for Solid State Physics, Dresden
Technical University, TU-Dresden, 01062 Dresden, Germany}
\author{M. Moroni}
\email{matteo.moroni01@ateneopv.it} \affiliation{Department of
Physics, University of Pavia-CNISM, I-27100 Pavia, Italy}
\author{L. Bossoni}
\affiliation{Department of Physics, University of Pavia-CNISM,
I-27100 Pavia, Italy} \affiliation{Kamerlingh Onnes Laboratory,
University of Leiden, 2300 RA Leiden, The Netherlands}
\author{S. Sanna}
\affiliation{Department of Physics, University of Pavia-CNISM,
I-27100 Pavia, Italy}
\author{R. Kappenberger} \affiliation{Leibniz-Institut f\"ur
Festk\"orper- und Werkstoffforschung (IFW) Dresden, 01171 Dresden,
Germany}
\author{S. Wurmehl} \affiliation{Leibniz-Institut f\"ur
Festk\"orper- und Werkstoffforschung (IFW) Dresden, 01171 Dresden,
Germany} \affiliation{Institute for Solid State Physics, Dresden
Technical University, TU-Dresden, 01062 Dresden, Germany}
\author{A.U.B. Wolter} \affiliation{Leibniz-Institut f\"ur
Festk\"orper- und Werkstoffforschung (IFW) Dresden, 01171 Dresden,
Germany} \affiliation{Institute for Solid State Physics, Dresden
Technical University, TU-Dresden, 01062 Dresden, Germany}
\author{M.A. Afrassa} \affiliation{Leibniz-Institut f\"ur
Festk\"orper- und Werkstoffforschung (IFW) Dresden, 01171 Dresden,
Germany}\affiliation{Addis Ababa University, College of Natural
Science, Addis Ababa, Ethiopia}
\author{ Y. Kobayashi}
\affiliation{Department of Physics, Division of Material Science,
Nagoya University, Nagoya 464-8602, Japan}
\author{M. Sato}
\affiliation{Research Center for Neutron Science and Technology,
CROSS, 162-1 Shirakata, Tokai 319-1106 Japan}
\author{B. B\"uchner} \affiliation{Leibniz-Institut f\"ur
Festk\"orper- und Werkstoffforschung (IFW) Dresden, 01171 Dresden,
Germany} \affiliation{Institute for Solid State Physics, Dresden
Technical University, TU-Dresden, 01062 Dresden, Germany}
\author{P. Carretta}
\affiliation{Department of Physics, University of Pavia-CNISM,
I-27100 Pavia, Italy}

%%%%%%%%%%%%%%%%%%%%%%%%%%%%%%%%%%%%%%%
\begin{abstract}
$^{19}$F NMR measurements in optimally electron-doped
La$_{1-y}$Y$_{y}$Fe$_{1-x}$Mn$_{x}$AsO$_{0.89}$F$_{0.11}$
superconductors are presented. In these materials the effect of Mn
doping on the superconducting phase is studied for two series of
compounds ($y= 0$ and $y= 0.2$) where the chemical pressure is
varied by substituting La with Y. In the $y=0.2$ series
superconductivity is suppressed for Mn contents an order of
magnitude larger than for the $y=0$ series. For both series a peak
in the $^{19}$F NMR nuclear spin-lattice relaxation rate $1/T_1$
emerges upon Mn doping and gets significantly enhanced on
approaching the quantum phase transition between the
superconducting and magnetic phases. $^{19}$F NMR linewidth
measurements show that for similar Mn contents magnetic
correlations are more pronounced in the $y=0$ series, at variance
with what one would expect for $\vec Q=(\pi/a,0)$  spin
correlations. These observations suggest that Mn doping tends to
reduce fluctuations at $\vec Q=(\pi/a,0)$ and to enhance other
low-frequency modes. The effect of this transfer of spectral
weight on the superconducting pairing is discussed along with the
charge localization induced by Mn.
\end{abstract}
%%%%%%%%%%%%%%%%%%%%%%%%%%%%%%%%%%%%%%%%


\maketitle

%%%%%%%%%%%%%%%%%%%%%%%%%%%%%%%%%%%%%%%
%%%%%%%    INTRODUCTION      %%%%%%%%%%
%%%%%%%%%%%%%%%%%%%%%%%%%%%%%%%%%%%%%%%
The introduction of impurities in superconducting materials is a
well known approach to test their stability for future
technological applications as well as to unravel their intrinsic
microscopic properties. In the cuprates the study of the staggered
spin configuration around isolated spin-less impurities, as Zn,
has allowed to determine how the electronic correlations evolve
throughout the phase diagram.\cite{Alloul} When a sizeable amount
of impurities is introduced they can no longer be treated as
independent local perturbations, the correlation among the
impurities themselves has to be considered and quantum transitions
to new phases may arise.\cite{Alloul1}

In iron-based superconductors several studies on the effect of
impurities have been carried out and it has soon emerged that the
behaviour may vary a lot depending on the family considered. If
one concentrates on the LnFeAsO$_{1-z}$F$_z$ (Ln1111) family, with
Ln a lanthanide ion, one notices that diamagnetic impurities
introduced by substituting Fe by Ru, cause a very weak effect both
on the magnetic ($z=0$)\cite{McGuire,Yiu1,Yiu2,Bonfa} and on the
superconducting
($z=0.11$)\cite{Sato1,Lee,Sato1B,SannaPRL,Sanna1su4} ground-state.
One has to introduce almost 60\% of Ru to quench either one of the
two phases.  On the other hand, if one considers the effect of
paramagnetic impurities as Mn a much stronger effect is observed.
In fact, in the optimally doped ($z\simeq 0.11$) Sm1111 the
superconducting transition temperature $T_c$ vanishes for a Mn
content around 8\% (see Fig. 1). Remarkably in  optimally
electron-doped LaFe$_{1-x}$Mn$_x$AsO$_{0.89}$F$_{0.11}$
superconductors $T_c$ drops to zero for $x$ as low as $0.2$\%
\cite{Sato2}, more than an order of magnitude less than for
Sm1111, and a quantum phase transition to a magnetic ground-state
is observed.\cite{FranziskaMn} The different behaviour of Sm1111
and La1111 against Mn impurities shows that by decreasing the
lanthanide ion size $T_c$ decreases more slowly with $x$  and the
system is driven away from a quantum critical point (QCP).
%%%%%%%%%%%%%%%%%%%%%%%%%%%%%%%%%%%%%%%%%%%%
\begin{figure}[h!]
\vspace{6.5cm} \includegraphics{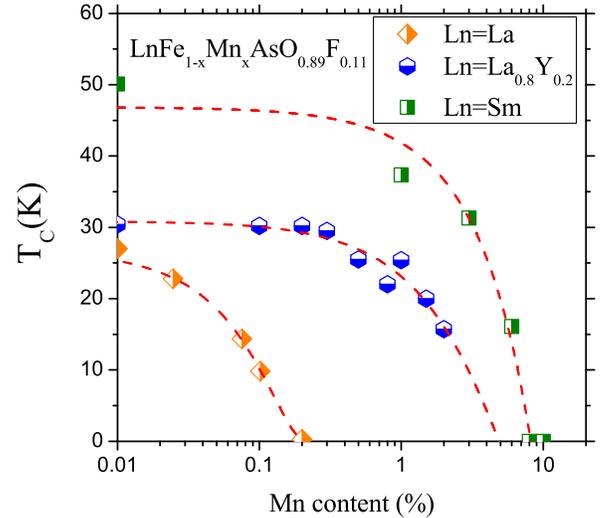} \caption{(Color online) Superconducting
phase diagram for
LaFe$_{1-x}$Mn$_{x}$AsO$_{0.89}$F$_{0.11}$,\cite{FranziskaMn}
La$_{0.8}$Y$_{0.2}$Fe$_{1-x}$Mn$_{x}$AsO$_{0.89}$F$_{0.11}$ and
SmFe$_{1-x}$Mn$_{x}$AsO$_{0.89}$F$_{0.11}$ \cite{Sato1,Sato1B}  vs
Mn content. The values of T$_c$  were determined from SQUID
magnetization measurements. Dashed lines are guides to the eye.}
\label{fig1}
\end{figure}
%%%%%%%%%%%%%%%%%%%%%%%%%%%%%%%%%%%%%%%%%%%%

The nature of the magnetic ground-state developing at high Mn
contents is still controversial.\cite{MGBA} In
Ba$_{0.5}$K$_{0.5}$(Fe$_{1-x}$Mn$_{x}$As)$_{2}$ superconductors
neutron scattering results suggested that Mn could modify the
magnetic wave-vector from $(\pi/a,0)$ to $(\pi/a,\pi/a)$ (square
lattice unit cell with Fe ions at the vertexes),\cite{Tucker}
leading to a weakening of $s_{\pm}-$wave pairing.\cite{Millis}
However, the absence of Ln1111 single crystals with a size
appropriate for neutron scattering experiments makes the
determination of the magnetic correlations developing upon Mn
doping rather difficult for this family of superconductors.

Since in Fe-based superconductors one of the most likely pairing
mechanisms involves magnetic excitations,\cite{Magpairing} it is
of major importance to investigate how the spin excitations evolve
in  optimally doped LaFe$_{1-x}$Mn$_x$AsO$_{0.89}$F$_{0.11}$
superconductors as the QCP is approached. From $^{75}$As nuclear
spin-lattice relaxation rate $1/T_1$ measurements it was
observed\cite{FranziskaMn} that when $T_c$ vanishes for $x_c\simeq
0.002$ the spin correlations follow the behaviour predicted for
strongly correlated electron systems close to a two-dimensional
(2D) antiferromagnetic (AF) QCP.\cite{Moriya} In this manuscript
it is shown that upon increasing the chemical pressure, by
partially substituting Y for La, $T_c$ decreases more slowly with
$x$ (Fig.1), mimicking the effect observed for Sm1111. This
indicates that the different behaviour of Sm1111 and La1111
against Mn impurities has to be associated with the larger
chemical pressure induced by the lanthanide ions on the FeAs
planes in the former case. From $^{19}$F NMR $1/T_1$ measurements
it is shown that besides the high frequency ($10^{12}- 10^{13}$
s$^{-1}$) dynamic probed by $^{75}$As nuclei, an additional very
low-frequency (MHz range) dynamic develops upon Mn doping and gets
progressively enhanced as the QCP is approached. Furthermore, it
is evidenced that if the system is driven away from the QCP by
partially substituting La with Y these low-frequency dynamic gets
significantly enhanced only at high Mn contents where
$T_c\rightarrow 0$. These results evidence that the disruption of
the superconducting phase coincides with the enhancement of
low-frequency fluctuations possibly competing with the ones
driving superconductivity.

NMR experiments were performed on
(La,Y)Fe$_{1-x}$Mn$_{x}$AsO$_{0.89}$F$_{0.11}$ polycrystalline
samples. Y for La substitution allows to vary the chemical
pressure without introducing paramagnetic lanthanide ions which
would significantly affect $^{19}$F NMR $1/T_1$.\cite{Prando} Two
series of samples were studied, the first one with no Y and with
Mn contents of $x=$ 0\%, 0.025\%, 0.075\%, 0.1\%, 0.2\% (referred
to as LaY0), while the second one with 20\% of Yttrium (LaY20
hereafter) and Mn content $x=$ 0\%, 0.3\%, 0.5\%, 10\%, 20\%. LaY0
samples were prepared as described in Ref. \onlinecite{Sato2},
while LaY20 as described in \onlinecite{SupplMat}. All the samples
were optimally electron doped with fluorine content around 11\%
.\cite{checkF} $T_c$ was determined by means of Superconducting
Quantum Interference Device (SQUID) zero field-cooled
magnetization measurements in a 10 Oe magnetic field. The diagram
of the superconducting phase, for both series of samples, as a
function of Mn content is shown in Fig.~\ref{fig1}. It is evident
that the introduction of 20\% of Yttrium stabilizes the
superconducting phase, leading to an increase of $T_c$ for $x=0$
\cite{Martinelli2009} as well as to a marked increase of $x_c$
from 0.2\% to about 4.5\%.

\begin{figure}[h!]
\vspace{10.8cm} \includegraphics{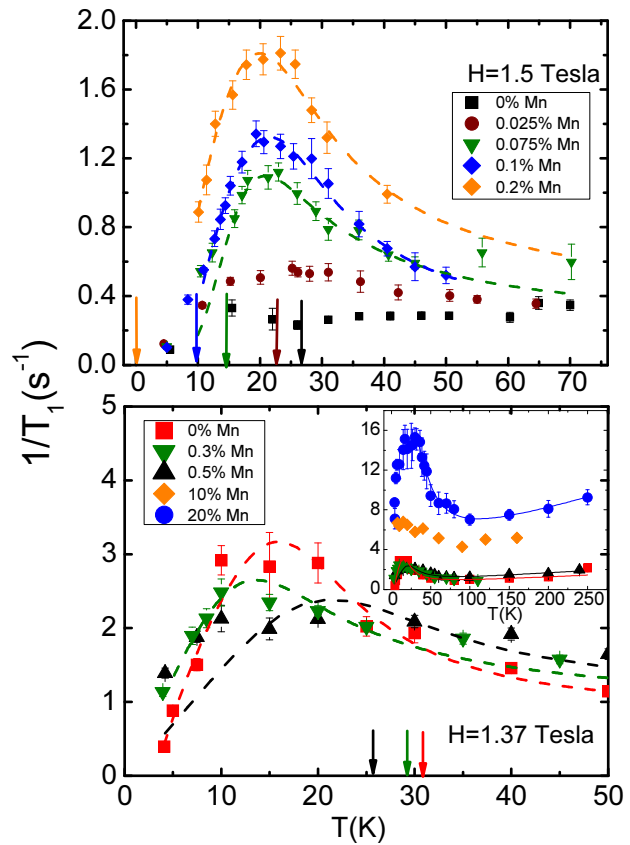} \caption{(Color online) Top: Temperature
dependence of $^{19}$F NMR 1/T$_1$ for LaY0 with $x$ up to 0.2\%.
Bottom: Temperature dependence of $^{19}$F NMR 1/T$_1$ in LaY20
for Mn contents up to $x =$0.5\%. In the inset the temperature
dependence is reported also for $x=10$\% and $x= 20$\%
non-superconducting samples. The dashed lines are fits of the data
according to Eq.3 in the text, while the arrows indicate the $T_c$
of the different samples (decreasing with increasing $x$).}
\label{fig2}
\end{figure}

$^{19}$F NMR measurements were performed at low magnetic fields,
$H\leq 1.5$ Tesla, by using standard radiofrequency pulse
sequences. The spin-lattice relaxation rate was estimated by
following the recovery of nuclear magnetization $M_z(\tau)$ after
a saturation recovery sequence. The recovery was fit according to
\begin{equation} \label{recovery}
M_z(\tau)=M_0[1-f\, e^{-(\tau/T_1)^\beta}]
\end{equation}
with M$_0$ the magnetization at equilibrium. The factor $f\simeq
1$ is introduced to account for incomplete saturation and $\beta$
is a stretching exponent which indicates a distribution of
$1/T_1$. The stretching exponent $\beta$ was found to be 1 for $T>
80$ K and decreased to about 0.5 at low temperature. The
distribution of relaxation rates originates from the presence of
different inequivalent Mn impurity configurations around $^{19}$F
nuclei.

The temperature dependence of $1/T_1$ for both sample series is
shown in Fig.~\ref{fig2}. Below 70 K $^{19}$F NMR $1/T_1$ is
characterized by a progressive increase upon decreasing the
temperature, by a pronounced maximum around 20 K, which can be
either below or above $T_c$ depending on the Y and Mn content (see
Fig.1) and eventually by a decrease at low temperature. Since, in
all samples, the increase starts well above $T_c$ those peaks
should not be associated with dynamics which develop in the
superconducting phase (e.g. vortex motions)\cite{BossoniVort} but
to normal state low-energy excitations. It should be remarked that
in the normal phase of Ln1111 iron-based superconductors without
impurities no marked peak in $1/T_1$ have ever been reported. Only
peaks in $1/T_1T$ have been observed,\cite{FranziskaBPP}
corresponding to small bumps in $1/T_1$. Here it is noticed that
those peaks are significantly enhanced by the presence of
impurities suggesting that Mn  tends to strengthen those
low-frequency dynamics which might already be present in the pure
compounds (see Fig. 2).

\begin{figure}[h!]
\vspace{5.8cm} \includegraphics{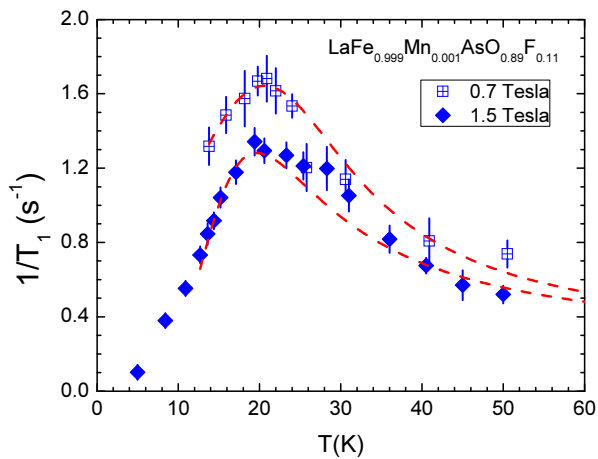} \caption{(Color online) $^{19}$F NMR
$1/T_1$ in the $x=0.1$\% LaY0 sample at two different magnetic
fields: 0.7 and 1.5 Tesla. Dashed lines are fits according to
Eq.3.} \label{fig4}
\end{figure}

By performing $1/T_1$ measurements at different magnetic fields
one observes that while the high temperature behaviour is only
weakly field dependent the magnitude of the peak around 20 K grows
by lowering the magnetic field (Fig. 3). This is exactly the
behaviour expected in the presence of dynamics approaching the
nuclear Larmor frequency $\omega_0$, namely in the MHz range. If
one assumes an exponential decay for the correlation function
describing the fluctuations with a characteristic time $\tau_c$,
then one can write\cite{BPP2}
\begin{equation} \label{eqbpp}
\dfrac{1}{T_1}= \gamma^2 <h^2_\perp>
\dfrac{\tau_c}{1+\omega^2_0\tau_c^2}
\end{equation}
where $\gamma$ is the nuclear gyromagnetic ratio and $<h^2_\perp>$
the mean square amplitude of the local field fluctuations
perpendicular to $\vec H$. In several disordered systems,
including cuprates,\cite{MHJ} the temperature dependence of the
correlation time is well accounted for by an Arrhenius law $\tau_c
(T) = \tau_0\exp(E_a/k_BT)$ with $E_a$ an energy barrier and
$\tau_0$ the correlation time at infinite temperature.
Nevertheless, a monodispersive behaviour cannot suitably describe
the broad peaks in $1/T_1$ and one rather has to consider a
distribution of correlation times associated with the non-uniform
distribution of Mn impurities. This corresponds to a distribution
of energy barriers which, for simplicity, was taken as squared
with a width $\Delta$ centered around $<E_a>$.\cite{Filibian2007}
In order to account for the high temperature behaviour of $1/T_1$
a linear Korringa term $\alpha T$,\cite{Slichter} characteristic
of metallic systems was introduced (see Fig.~\ref{fig2}). Then
$1/T_1$ can be described by the expression:\cite{Filibian2007}
\begin{eqnarray} \label{T1eq}
\dfrac{1}{T_1}= \frac{\gamma^2 <h^2_\perp> T}{2\omega_0
\Delta}\biggl[ atan\biggl( \omega_0\tau_0
e^{(<E_a>+\Delta)/T}\biggr) -\nonumber \\ atan\biggl(
\omega_0\tau_0 e^{(<E_a>-\Delta)/T}\biggr) \biggr] + \alpha T
\end{eqnarray}
By fitting the data of the superconducting samples ($x< 0.2$\% for
LaY0 and $x\leq 0.5$\% for LaY20) one notices that for LaY0 spin
correlations yield a significant increase in the width of the
distribution on approaching the crossover between the
superconducting and magnetic phases (Fig. 4). On the other hand,
for the LaY20, within the uncertainty of the fit parameters, there
is no evidence for a neat increase of $\Delta$ in the same doping
range (Fig. 4). In other terms, in the LaY20 family the dynamic
does not vary significantly upon increasing the Mn content up to
$x= 0.5$\%, suggesting that the collective coupling is still weak.

\begin{figure}[h!]
\vspace{5cm} \includegraphics{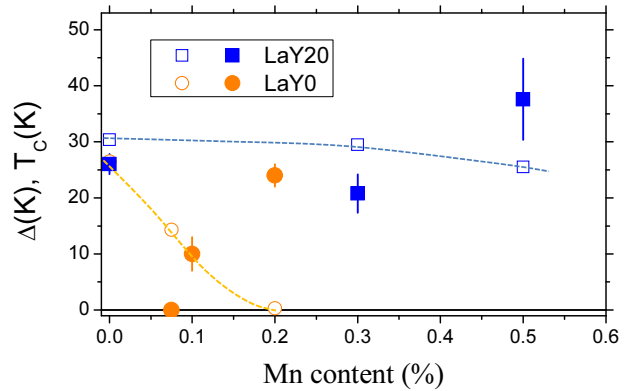} \caption{(Color online) The width of the
distribution of energy barriers $\Delta$ (closed symbols) and
$T_c$ (open symbols) are reported for the LaY0 and LaY20 samples.
$\tau_0=3\pm 2 \times 10^{-10}$~s for both families, while
$<E_a>=47$~K fixed for LaY0 and $<E_a>=33\pm 8$~K for LaY20
compounds.} \label{BPPParam}
\end{figure}

\begin{figure}[h!]
\vspace{5.5cm} \includegraphics{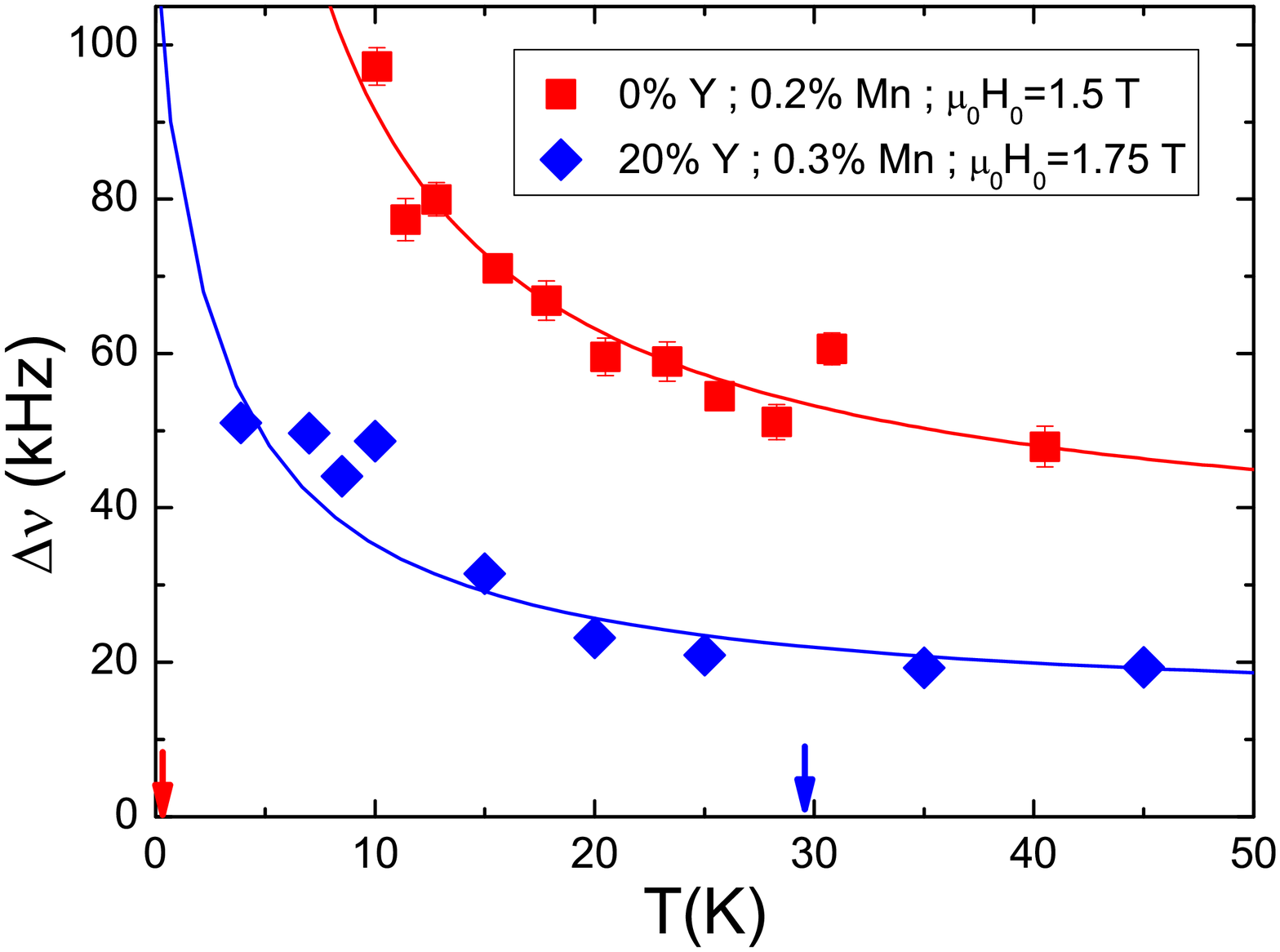} \caption{(Color online) $^{19}$F NMR line
full width at half maximum $\Delta\nu$ in the $x=0.2$\% LaY20
sample (blue diamonds) and in the LaY0 $x=0.3$\% sample (red
squares). Solid lines are best fits according to a Curie-Weiss law
(see text), while the arrows indicate the $T_c$ of the two
samples.} \label{fig3}
\end{figure}

Moreover, one may notice (Fig. 2) that for $x\leq 0.5$\%, in the
LaY0 series the peak in $1/T_1$ grows significantly with Mn
doping, while in the LaY20 series it remains practically
unchanged. This evidences that $1/T_1$ increases progressively as
$T_c\rightarrow 0$, namely the strength of the local spin
susceptibility in the FeAs plane gets enhanced due to the
proximity to the QCP. In other terms, for similar Mn contents the
spin correlations get weaker as the chemical pressure is increased
by Y doping. Further support in this respect is provided by the
temperature dependence of $^{19}$F NMR linewidth $\Delta\nu$ which
is directly related to the amplitude of the staggered
magnetization developing around the impurity.\cite{Bobroff} As
shown in Fig. \ref{fig3} for similar Mn doping the $^{19}$F NMR
linewidth is unambiguously larger in the sample without Y. The
data in Fig. \ref{fig3} can be fitted with a Curie-Weiss law
$\Delta\nu= (\Delta\nu)_0+ C/(T+\Theta)$. $(\Delta\nu)_0\simeq 13$
kHz is the temperature-independent linewidth due to nuclear
dipole-dipole interaction which is assumed similar for both
samples, since it is determined by $^{19}$F-$^{19}$F dipolar
coupling which remains practically unchanged (the F content and
the lattice parameters do not vary significantly between the two
samples). The fit of the data shows that $C$ increases by a factor
3 and that $\Theta$ increases from about 3 K to 11 K between
$x=0.3$\% LaY20 and $x=0.2$\% LaY0 sample.

The observation that the magnetic correlations get depressed when
La is substituted by a smaller lanthanide ion can in principle be
associated with a decrease of the ratio $U/t$ between Coulomb
repulsion and hopping integral due to the increase in the chemical
pressure. However, for stripe collinear order ($\vec Q=(\pi/a,0)$
or $(0,\pi/a)$) theoretical works\cite{Band,Giovannetti} suggest
that in Ln1111 the magnetic order parameter should get enhanced on
decreasing the Ln size or equivalently increasing As $z/c$
coordinate, exactly the opposite of what is found here. It should
also be remarked that the behaviour found upon Mn-doping is the
contrary of that observed in Ru-substituted Fe-based
superconductors where the magnetic order is stabilized by
decreasing the size of the lanthanide.\cite{Sanna1su4} Hence, it
is likely that upon increasing $x$ magnetic correlations different
from the stripe ones develop. Giovannetti et
al.,\cite{Giovannetti}, in the framework of Landau free energy
calculations, showed that around optimal electron doping the
energy difference between the stripe and orthomagnetic phases,
with a $\pi/2$ rotation of the spins, gets reduced. Hence, it
might be possible that the introduction of Mn impurities could
stabilize the latter type of order.

More recently Gastiasoro and Andersen \cite{Andersen} have
considered the cooperative behaviour of paramagnetic impurities
introduced in the FeAs planes of  Fe-based superconductors,
coupled via an RKKY interaction. They pointed out that upon
increasing the Kondo-like coupling between the localized impurity
and the itinerant electrons, N\'eel ($\vec Q=(\pi/a,\pi/a)$)
correlations would arise and the amplitude of collinear stripe
modes decrease. However, even when the coupling gets significant
and N\'eel fluctuations enhanced the stripe spin correlations
would still survive. In a real space description their results
imply the development of N\'eel type correlations in small islands
around the impurity and stripe spin arrangement at large distances
from the impurity. Even if from our $^{19}$F NMR spectra one
cannot check the validity of this model, this theoretical approach
is able to explain both the weakening of the superconductivity
\cite{Millis} and the onset of a novel magnetic phase upon Mn
doping.\cite{FranziskaMn} In such a scenario the peaks in $1/T_1$
should be associated with the freezing of the spin fluctuations
around Mn impurities which get more and more correlated as the QCP
is approached. The theoretical model by Gastiasoro and Andersen
\cite{Andersen} also allows to make an analogy between heavy
fermion physics and the one achieved by doping Fe-based
superconductors impurities. In this respect we recall that,
similarly to heavy fermions, at the QCP there is a charge
localization \cite{Sato2} suggesting a divergence of the electron
effective mass. Hence, one should actually consider two possible
concomitant effects which depress superconductivity: loss of spin
excitations causing the pairing and/or charge localization. Once
more, we remark that the behaviour achieved by Mn-doping is quite
different from the one observed in Ln1111 superconductors doped
with Ru spin-less impurities, where even at very high doping
levels ($\simeq 50$\%) the system remains metallic.\cite{Tropeano}

An alternative explanation for the growth of low-frequancy spin
fluctuations in Mn-doped Ln1111 relies on the presence of nematic
fluctuations. In this respect it is interesting to observe that
even nominally pure samples do show a small bump in $1/T_1$
\cite{FranziskaBPP,BossoniT2,Dioguardi} in the same temperature
range where the peak in $^{19}$F NMR $1/T_1$ arises. The same
low-frequency dynamic was found to affect the NMR transverse
relaxation rate in Ba(Fe$_{1-x}$Rh$_{x}$)$_{2}$As$_{2}$ and was
tentatively associated with nematic fluctuations, possibly
involving charge stripes.\cite{BossoniT2,Dioguardi} Although there
is no neat evidence for these type of dynamics here, one may be
tempted to relate the energy barrier probed by $1/T_1$ to the one
separating the degenerate nematic phases.\cite{Nematic} In this
framework the enhancement of the low-frequency dynamics could be
associated with the pinning of those fluctuations driven by Mn.

In conclusion, the increase in the chemical pressure driven by Y
for La substitution in
La$_{1-y}$Y$_{y}$Fe$_{1-x}$Mn$_{x}$AsO$_{0.89}$F$_{0.11}$ is found
to lead to a less effective suppression of the superconducting
ground-state by Mn doping. $^{19}$F NMR $1/T_1$ measurements
exhibit a low-temperature peak which indicates the onset of very
low-frequency dynamics with an amplitude directly related to the
proximity of the compound to the QCP between superconducting and
magnetic phases. Based on recent theoretical works, this behaviour
could be ascribed to the enhancement of spin correlations
different from stripe ones, suggesting that $T_c$ is depressed by
the decrease in the spin fluctuations around $(\pi/a,0)$, which
are widely believed to mediate the pairing, or by the localization
effect in the region close to the metal-insulator boundary.

We would like to acknowledge useful discussion with Brian
Andersen, Maria N. Gastiasoro and J. Lorenzana. This work was
supported by MIUR-PRIN2012 project No. 2012X3YFZ2, by the DFG
through the SPP1458 in project BU887/15-1 and SFB1143 and by the
Emmy-Noether program (Grant No. WU595/3-1). We acknowledge R.
Wachtel, S. Müller-Litvani, and G. Kreutzer for technical support.

%%%%%%%%%%%%%%%%%%%%%%%%%%%%%%%%%%%%%%%


\begin{thebibliography}{99}
\bibitem{checkF} For the LaY0 samples no variation
of the intensity of the $^{19}$F-NMR resonance line was found
within the error bars, confirming that the intrinsic F content
does not differ among the samples within $\pm 0.005$.
\end{thebibliography}
\end{document}